\documentstyle[aps,preprint,prc,12pt]{revtex}
\begin{document}
%\baselineskip 24pt
%\baselineskip 20pt
%\vspace{0.1 truecm}

\begin{center}

{\Large \bf  Isotopic composition of fragments in multifragmentation of very 
large nuclear systems: effects of the chemical equilibrium} 
\end{center}
%\vspace{0.2 truecm}

\begin{center}
{\large P.M.Milazzo$^{1}$, A.S.Botvina$^{2}$, G.Vannini$^{3}$, M.Bruno$^{3}$, 
N.Colonna$^{4}$, M.D'Agostino$^{3}$, F.Gramegna$^{5}$, I.Iori$^{6}$, 
G.V.Margagliotti$^{1}$, P.F.Mastinu$^{5}$, A.Moroni$^{6}$, R.Rui$^{1}$
}
\end{center}

%\vspace{0.2 truecm}

{\it 
$^{1}$Dipartimento di Fisica and INFN, 34127 Trieste, Italy

$^{2}$Gesellschaft fur Schwerionenforschung, Darmstadt, Germany, 
and Institute for Nuclear Research, Russian Academy of Science, 
117312 Moscow, Russia

$^{3}$Dipartimento di Fisica and INFN, 40126 Bologna, Italy

$^{4}$INFN, 70126 Bari, Italy

$^{5}$INFN, 35020 Laboratori Nazionali di Legnaro, Italy

$^{6}$Dipartimento di Fisica and INFN, 20133 Milano, Italy
%\vspace{0.1 truecm}
\begin{abstract}
Studies on the isospin of fragments resulting from the disassembly 
of highly excited large thermal-like nuclear emitting sources, formed in the 
$^{197}$Au+$^{197}$Au reaction at 35 MeV/nucleon beam energy, are presented. 
Two different decay systems (the 
quasiprojectile formed in midperipheral reactions and the unique source 
coming from the incomplete fusion of projectile and target in the most central 
collisions) were considered; these emitting sources have the same initial N/Z 
ratio and excitation energy (E$^* \simeq$ 5--6 MeV/nucleon), but different 
size. 
Their charge yields and isotopic content of the fragments show different 
distributions. 
It is observed that the neutron content of intermediate mass fragments 
increases with the size of the source. These evidences are 
consistent with chemical equilibrium reached in the systems.
%[In this sense, to increase the size 
%of the emitting source will produce at the de-excitation stage a formation of 
%a number of fragments higher than that expected from the simple source size 
%factor. Then, even if some input parameters are fixed (E$^*$ and N/Z), the 
%partition function present different behaviours with the size of the system. ]
This fact is confirmed by 
the analysis with the statistical multifragmentation model. 
\end{abstract}

%\vspace{0.2 truecm}

\hspace{0.5 truecm}PACS numbers: 25.70Pq, 25.70-z, 24.60-k

\newpage

%\section{Introduction}
\rm
The study of heavy-ion collisions at intermediate energies 
(10$\leq$E$\leq$100 MeV/nucleon) is a useful tool to investigate 
the mechanisms of fragment production in highly excited nuclear systems. 
In this energy regime multifragmentation appears as one of the main 
de-excitation channel \cite{hirschegg}, which can be considered 
\cite{exp,pochod} as a manifestation of a liquid-gas type phase transition in 
finite nuclear systems. This gives an access to the nuclear equation of state. 
It has been shown that statistical models \cite{PR95,gross} happen to 
be very effective in the reproduction of main characteristics of the fragment 
production such as charge distributions and fragment correlations. 

Nowadays, with the advent of radioactive beam facilities, the influence 
of the isospin degree of freedom is strongly addressed and experimental 
information on the isotopic content of emitted fragments represents a 
meaningful starting point in order to get a deep understanding of either 
the de-excitation mechanisms or the nuclear matter properties. 

Theoretical calculations predict that, in the fragmentation of asymmetric 
nuclear matter, the isospin composition of the "liquid" phase (usually 
associated with intermediate mass fragments (IMF) and heavy residues) and the 
"gas" phase (light charged particles (LCP) and nucleons) depends on many 
factors. 
In particular, calculations by M\"uller and Serot \cite{MS} showed that, 
for very neutron-rich systems, there may exist a distribution of the excited 
nuclear matter into a neutron-rich gas and a more symmetric liquid. 
However, Lee and Mekjian \cite{Lee2001} have pointed out that the Coulomb 
and surface effects, which are important for finite systems, may moderate 
the neutron enrichment of the gas phase by producing more free protons. 
The Statistical Multifragmentation Model (SMM) calculations \cite{botmis01} 
have shown that the neutron content of the IMF can increase in the region of 
the phase transition, in agreement with the experimental observations 
\cite{milaz00}.

In this situation only new experimental data could solve the problem of 
isospin composition of the gas and liquid phases. 
Indeed, a variety of experiments can be found in literature. 
In Refs.~\cite{veselski,martin} the difference of the mean N/Z ratio of the 
LCP and of the IMF 
has been interpreted as the separation of gas and liquid phases. 
The authors of Ref.~\cite{xu00} considered the presence of neutron rich 
LPC and light IMF as an evidence of the neutron enrichment of the gas phase. 
%Quite neutron deficient residues have also been observed in incomplete fusion 
%studies \cite{hanold}. 
The average N/Z ratio of fragments emitted from excited nuclear systems is 
seen to vary with the excitation energy and the N/Z ratio of the system. 
In particular in the work of 
Ramakrishnan et al. \cite{rama}, where for different reaction the beam energy 
and the mass of the system were kept constant, it has been shown that the 
IMF's isospin present a linear dependence on the N/Z ratio of the system. 
Moreover, it has been observed that the neutron content of 
the emitted IMF depends on the excitation energy of the (fixed size) 
sources \cite{milaz00}, increasing as the excitation energy increase. 

In this paper we present experimental data extracted from two emitting sources 
with approximately the same N/Z ratio and excitation energy, but with 
different size. 
Since existing experimental data cover the study of the N/Z ratio of 
fragments as a function of the excitation energy and N/Z ratio of the 
emitting source, 
this analysis represent a complementary point to fill the experimental 
picture on the effects induced by the isospin on the decay process. 

It is also important to note that these new data provide information about 
isospin of fragments produced in decay of the largest nuclear systems under 
investigation up to now (A$>$300). This allows for more reliable extrapolations
for the case of nuclear matter, as well as for the astrophysical 
applications in supernovae explosions and neutron stars. 

%In Sect.II a brief description of the experimental conditions is 
%given; then the prescriptions used, to well identify the emitting systems and 
%to extract their characteristics, are described. In Sect.III the experimental 
%results are presented. In Sect.IV the interpretation of the measured 
%observables is discussed, then the conclusions are drawn in Sect.V. 

%\section{The experiment}
The reaction Au+Au at 35 MeV/nucleon was studied in experiments
performed at the National Superconducting K1200 Cyclotron
Laboratory of the Michigan State University. Light charged particles
and fragments with charge up to Z=20 were detected at
23$^{\circ}<\theta_{lab}<$160$^{\circ}$ by the phoswich detectors of
the MSU $Miniball$ hodoscope \cite{mb}. 
%The charge identification thresholds were about 2, 
%3, 4 MeV/nucleon in the $Miniball$ for Z= 3, 10, 18, respectively. 
The $MULTICS$ array \cite{mcs} covered the 
angular range 3$^{\circ}<\theta_{lab}<$23$^{\circ}$ and allowed for a 
charge discrimination up to Z=83 and for good mass discrimination for 
Z=1-6 isotopes. 
%The identification thresholds in the $MULTICS$ array were about 1.5
%MeV/nucleon for charge identification and about 10 MeV/nucleon for 
%mass identification. 
%The MULTICS array consisted of 48 telescopes, each of which was composed of an
%Ionization Chamber (IC), a Silicon position-sensitive detector (Si) 
%and a CsI crystal. Typical energy resolutions 
%were 2\%, 1\% and 5\% for IC, Si and CsI, respectively. 
The geometric acceptance of the combined array was
greater than 87\% of 4$\pi$. 

A brief summary of the results published in so far will follow. 
The selection of the impact parameter $\hat b$ is based on the 
number N$_c$ of charged particles detected \cite{nc}: 
$$\hat b = b/b_{max} = 
\left(\int_{N_c}^{+\infty} P(N'_c) dN'_c \right)^{1/2}.$$ 
where P(N$_c$) is the charged particle probability distribution and 
$\pi\cdot b_{max}^2$ is the measured reaction cross section for N$_c\ge$3. 

Through the analysis of energy and angular distributions of the experimental
data, it was possible to identify events originating either
from disassembly of a unique source formed in central collisions, 
or from decay of the quasiprojectile (QP) in peripheral and mid-peripheral
collisions \cite{milazzo98}. 
The excitation energy of the fragment sources was calculated both via the
calorimetric method and through comparison with model calculations 
\cite{milazzo98,dagostino96}. 
Extra kinetic energy, such as radial flow or rotational motion, does not
contribute appreciably to the excitation energy. This results from the
analysis of charge, angular and kinetic energy distributions of the emitted
fragments, as well as from the study of event by event charge partition 
\cite{milazzo98,dagostino96,dagostino99,deses}. 
Finally, nuclear temperatures were determined using the technique of the
double ratio of isotope yields \cite{albergo}.

The analysis determined that the thermal characteristics (Excitation energy 
E$^*$ and isotope temperature T$_{iso}$) of the QP, formed 
at 0.6$<\hat b<$0.7 (E$^*$=5.5$\pm$0.6 MeV/nucleon, T$_{iso}$=4.3$\pm$0.2 MeV),
and of the unique source (E$^*$=5.5$\pm$0.5 MeV/nucleon, 
T$_{iso}$=4.3$\pm$0.4 MeV) are remarkably similar \cite{milazzo98}.  
%The fact that emitting sources formed in midperipheral and central collisions 
%present similar values for temperature and excitation energy could be due to 
%the fact that in the central collisions a small amount of energy (less than 
%1 MeV/nucleon) might be gone in some collective motion (e.g. radial flow) and 
%some energy is released during the preequilibrium emission, 
%see e.g. \cite{PR95,deses}. 

%\section{Experimental results}
It is therefore important to analyse the whole picture 
of fragment production in these reactions. 
%For this purpose we remind that 
%the fragment charge distributions, their correlations and energy 
%characteristics were already presented in 
%Refs.\cite{milazzo98,dagostino96,dagostino99} 
%and analysed there within a statistical approach. 
The aim of this paper is to provide new results
by comparing IMF production between emitting sources of similar excitation
energies (E$^*\simeq$5--6 MeV/nucleon) 
and N/Z ratio (118/79=1.49), 
but of different size (Z=79 for the midperipheral QP, Z=126 for 
the central unique source (CUS) \cite{dagostino96}). 
In particular we will focus our attention to the isospin of emitted 
fragments.

Charge yields of fragments obtained for the Au+Au reactions at 35 MeV/nucleon
are presented in Fig.1. The heaviest source (CUS) decays emitting fragments
lighter than the QP, and in a number larger than what expected by a simple
scaling factor. In other words, the CUS undergoes a stronger disintegration
at the de-excitation stage.

The different partition in the two systems affects the number of
neutrons in the IMF. Fig.2 shows the isotope production yields for different
Z values. The IMF emitted from the CUS are more neutron rich than those from
the QP. This effect is enhanced when plotting the relative yield ratio vs
the neutron excess (see Fig.3a): the ratio between the CUS and QP yields
increases with (N-Z). In Fig.3b we present the ratio of relative yields of
neutron-rich to neutron-poor isotopes at fixed Z values, for both CUS and QP.
It appears that the CUS emits preferentially the more neutron rich fragments.
The average $<$N$>$/Z value of each atomic specie versus 
its charge Z is presented in Fig.4. 

One can clearly see from all the figures that the disassembly of the CUS 
system 
into a relatively larger number of fragments leads to production of more 
neutron rich IMF and LCP. In the previous study it was experimentally 
demonstrated that fixing the size of the source, but varying its excitation 
energy, the increase in the multiplicity of emitted fragments is accompanied 
to more neutron rich IMF \cite{milaz00}. 
In the present case we observe that this effect is even more pronounced. 

%\section{Interpretation of the data}
These results are consistent with the statistical picture of disintegration 
of finite nuclear systems. 
For example, 
the statistical multifragmentation model (SMM), see e.g. \cite{PR95}, 
is based upon the assumption of statistical equilibrium at a 
low-density freeze-out stage. 
%We consider all break-up channels (partitions) composed of nucleons and excited
%fragments taking into account mass, charge, momentum and energy conservation. 
%The statistical weight of decay channel $j$ is given by 
%$W_{j} \propto exp~S_{j}$, where $S_{j}$ is the entropy of 
%the system in channel $j$ depending on excitation energy $E_s^{*}$, 
%mass number $A_s$, charge $Z_s$, density $\rho_s$ and other parameters 
%of the source. 
%Light fragments with mass number $A\leq 4$ are considered as stable 
%particles ("nuclear gas") with only translational degrees of freedom; 
%fragments with $A > 4$ are treated as heated nuclear liquid  drops. 
%The Coulomb interaction between the fragments can 
%be treated either in the Wigner--Seitz approximation or calculated directly 
%for each spatial configuration of primary fragments in the freeze-out volume. 
Different break-up partitions are sampled, according to their statistical 
weights, in the phase space. After break-up of the nuclear source 
the fragments propagate independently in their mutual Coulomb field and 
undergo secondary decays. The de-excitation of the hot primary fragments 
proceeds via evaporation, fission or Fermi-break-up \cite{botvina87}.
SMM is very successful in reproducing experimental data 
concerning both peripheral and central nucleus-nucleus collisions 
\cite{botvina95,scharenberg,williams}. In particular a very good agreement 
between experimental data and SMM calculations was found in the study of 
the Au+Au 35 MeV/nucleon reaction \cite{milazzo98,dagostino96,dagostino99}. 
Here we use a new version of SMM based on the generation of a Markov 
chain of partitions \cite{botmis01}. This version keeps its full reliability 
concerning the charge distribution predictions and allows for taking into 
account all effects influencing the isotope content of the produced fragments.

%In the present analysis we don't try to reach the best agreement with the 
%data and don't use the filtering procedure, which, however, as was shown in 
%\cite{milazzo98,dagostino96}, change the results only slightly and practically 
%do not influence the relative isotope yields. 
In this paper, by basing on this statistical model we aim to single out 
the main physical effects responsible for the observed trend. 
Even though some assumptions of the model, such as a fixed freeze-out volume or 
non-overlapping fragments, are approximations of the real conditions, and some
parameters (flow energy, fragment's level density) are not clearly defined, 
these uncertainties influence the fragment charge distributions mainly and 
can be accounted for by finding the source distribution with the 
well elaborated technique \cite{deses}. 
When the charge distributions are fixed, the main features of the 
studied isotope distributions are driven by the chemical equilibrium effects 
and the binding energy of the fragments.

The events generated by SMM were filtered to take into account the 
experimental efficiency. 
In the Figures 1, 2 and 4 the comparisons between experimental data and  
SMM predictions is shown. 
The following set of parameters gives a good agreement between experimental 
data and calculations (see also \cite{milazzo98,dagostino96,deses}): 
an excitation energy $E_s^{*}$=5.5$\pm$0.6 MeV/nucleon; 
$A_s$=197, $Z_s$=79, $\rho_s=\frac{1} {3} 
\rho_0$, for the QP; $A_s$=315, $Z_s$=126, $\rho_s=\frac{1}{6} \rho_0$ 
for the CUS ($\rho_0$=0.15~$fm^{-3}$ is the normal nuclear density). 
The smaller density in the central collisions is consistent with an 
additional expansion caused by the flow development. The microcanonical 
temperatures obtained in both cases are also quite similar: 
$T_{micr} \approx$ 5.4~MeV for the central case and $\approx$ 5.5~MeV for the 
peripheral one. 

Dynamical calculations \cite{botvina95,botlarmis,tan2001} 
predict that N/Z ratios of the emitting sources must have values close to 
that of the initial system. In particular an analysis dedicated to the study 
of 50 MeV/nucleon central collisions of $^{112}$Sn+$^{112,124}$Sn 
\cite{tan2001} has shown that the best agreement between 
calculations and experimental data claims for N/Z values of the source 
close to that of the starting system (differences are lower than 3\%). 
Also the dynamical study of $^{129}$Xe+$^{197}$Au central and peripheral 
collisions at 50 MeV/nucleon has shown that the N/Z ratio of the 
thermal sources does not change essentially (differences are within 2\%) 
from the initial one \cite{botlarmis}. 
Results presented in Ref.\cite{botvina95} and concerning the 
study of peripheral Au+Cu reaction come to the same conclusions. 
Generally, since the considered emitting sources are large in size, 
fluctuations in their N/Z ratio value should be strongly reduced. 
Therefore, the assumption of conservation of the N/Z ratio after 
the dynamical stage, seems quite realistic. We note, that our following 
interpretation will be valid even under less strict assumptions, namely, 
when the N/Z ratio changes in the same way in the both cases, or it 
becomes smaller for the CUS.  
%Moreover, one could expect a lower N/Z value for the CUS, 
%since the higher probability to evaporate neutrons, while from the 
%experimental point of view, their emitted fragments are more neutron rich. 

The reason of why the nuclear sources at practically the same temperature 
and excitation energy produce so different fragment charge distributions 
(Fig.1), can be found in the larger Coulomb energy in the CUS case, and in 
the different dynamics of formation of the CUS leading to a flow and, as 
consequence, to smaller freeze-out densities. 

One can see also from the Figures that the calculations reproduce the trend 
of increasing neutron content of produced fragments with disintegration of the 
nuclei into smaller pieces. This trend can be explained as follows: 
if the chemical equilibrium is 
established the big fragments have larger N/Z ratio than small ones. 
After disintegration of the big fragments, independently if caused by 
increasing excitation energy (or temperature) of a thermal source as in 
\cite{milaz00} or by a size (Coulomb) effect, the neutrons of the big 
fragments are accumulated mainly in the small secondary fragments and not in 
free neutron gas. 
The SMM calculations predict that the mean number of primary free neutron 
in the freeze-out volume increases only from 0.98 for the peripheral source 
to 1.79 for the central source. 
This gives rise to the N/Z ratio of the IMF 
and LCP, which is preserved after the secondary de-excitation. 

Even if the model well explains the observed effects, the experimental rise in 
neutron content is slightly more pronounced. One can speculate, 
that the initial CUS could have slightly larger N/Z 
ratio than the peripheral source. Our calculations show that a slight 
decrease of this ratio to N/Z=1.43 without changing other SMM parameters, 
for the QP, would be sufficient to explain this disagreement. 
However, as we pointed above the difference (if any) of this ratio 
of the sources should not be significant. 
For this reason, we believe that the predicted redistribution of 
neutrons from heavy fragments to light ones 
with disintegration of a nuclear system is the natural mechanism for 
explanation of the data. 

%\section{Conclusion}
In summary in 
the study of the $^{197}$Au+$^{197}$Au 35 MeV/nucleon reaction it was 
possible to well identify two different emitting sources with the same N/Z 
ratio, excitation energy and temperature (E$^* \simeq$5--6 MeV/nucleon; 
T$_{iso} \simeq$ 4.3 MeV from the experimental measurements, 
T$_{micr} \simeq$ 5.4--5.5 MeV from the SMM predictions), but different size: 
the QP formed in midperipheral 
reactions and the CUS coming from the incomplete fusion of projectile and 
target in the most central collisions. 
Charge yields and isotopic content of the fragments show different 
distributions in the two cases. The CUS decays emitting lighter fragments than 
the QP and this fact reflects on the multiplicity of fragments from CUS, which 
is higher than what expected scaling for the size factor the QP multiplicity. 
Thus, even if some parameters are fixed (E$^*$, T, N/Z) the partition of 
the system depends on the size and the density 
which determine the Coulomb energy in the freeze-out. 
Moreover, the neutron content of light charged particles and intermediate 
mass fragments increases with the size of the source. 

%Several analyses of the multifragmentation process already established 
%many of its features. 
%It is well known that 
Primary fragments produced in the freeze-out are hot 
and decay afterwards by emitting mainly neutrons and light charged particles. 
Therefore, it is natural to connect the behaviour of the experimental 
isotope yield as a function of source size 
with the similar evolution of the N/Z ratio of 
the corresponding hot fragments. This conclusion is here supported by the SMM 
calculations which reproduce the fragment production in a reasonably good way. 
The data here presented, obtained both in central and peripheral 
\cite{milaz00} collisions, 
indicate that the neutron content of the fragments, produced by the 
decay of thermal-like systems, increases with the multiplicity of emitted IMF.
%Presently, we do not see a chance to explain these data within 
%the scenario obtained in some 
%models \cite{MS}, which suggests the separation into neutron rich 
%gas and isospin symmetric "liquid" fragments during the nuclear phase 
%transition. 
The data are consistent with hypothesis of thermal and 
chemical equilibration in finite nuclear systems \cite{botmis01}, which 
leads to production of IMF with large neutron content. 
This justifies 
the method of extracting temperatures through isotope thermometers 
\cite{milazzo98} and the thermodynamical description of these reactions. 

%\centerline{\bf Acknowledgements}

The authors are indebted to I.N.Mishustin for his contribution to 
the development of the new version of SMM.
The authors would like also thanks J.D.Dinius, S.Gaff, C.K.Gelbke, 
T.Glasmacher, M.J.Huang, W.J.Lynch, C.P.Montoya, M.B.Tsang and H.Xi, for 
their collaboration during measurements and data analysis.
One of the author (A.S.B.) appreciate warm hospitality of INFN (sezione di 
Trieste) where a part of this work was done.

\vspace{0.3 truecm}

{\large \bf  {Figure captions}}\\
%{\bf Fig.1:} {Mean elemental event multiplicity N(Z): solid and open points 
%are for the QP and the CUS, respectively.}
%\vspace{0.5 cm}

{\bf Fig.1:} {Charge yields of fragments obtained for the Au+Au reactions 
at 35 MeV/nucleon \cite{milazzo98,dagostino96}. 
Circles represent the experimental data: the solid ones are for the 
QP systems, the open ones are for the CUS. 
Solid and dashed lines are SMM calculations for 
peripheral and central cases, respectively.}
\vspace{0.5 cm}

{\bf Fig.2:} {Relative yields of different isotopes for fragments with charges 
from Z=1 to Z=6. 
Notations are as in Fig.1. }
\vspace{0.5 cm}

{\bf Fig.3:} {(a) ratio between the CUS and QP yields for each isotope as a 
function of N-Z. (b) ratio of relative yields of neutron-rich to 
neutron-poor isotopes at fixed Z values, for both cases (solid point are for 
the QP, open points for the CUS).}
\vspace{0.5 truecm}

%{\bf Fig.3:} {Mean neutron--to--proton (N/Z) ratio 
%of the produced fragments. Notations are as in Fig.~1.}
%\vspace{0.5 truecm}

{\bf Fig.4:} {Mean neutron--to--proton ($<N>/Z$) ratio 
of the produced fragments. 
Notations are as in Fig.1. }
%Circles represent the experimental data: 
%solid points for the QP, open points for the CUS. Solid and dashed lines are 
%the SMM calculations for 
%peripheral and central cases, respectively.}
\vspace{0.5 truecm}

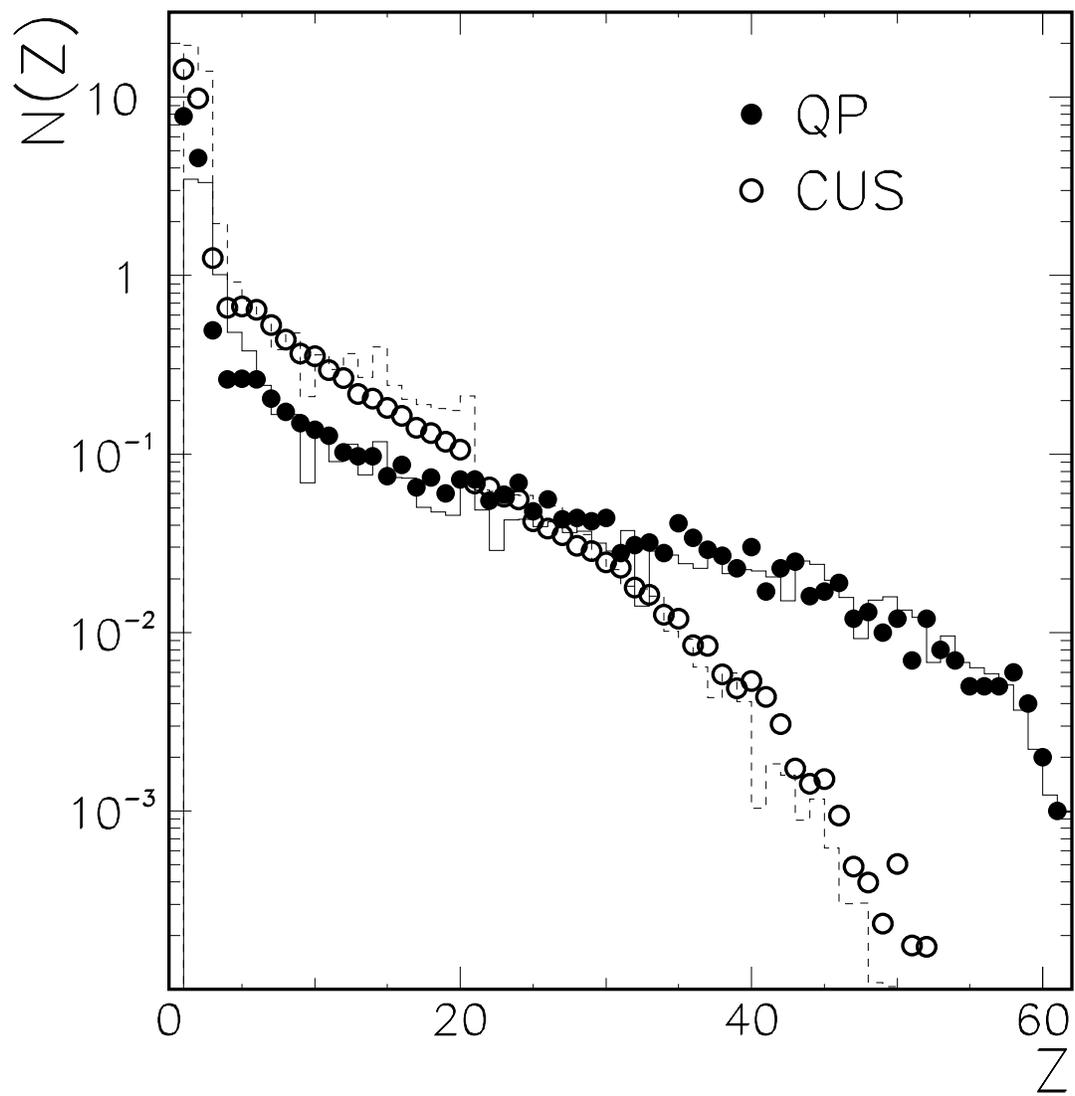
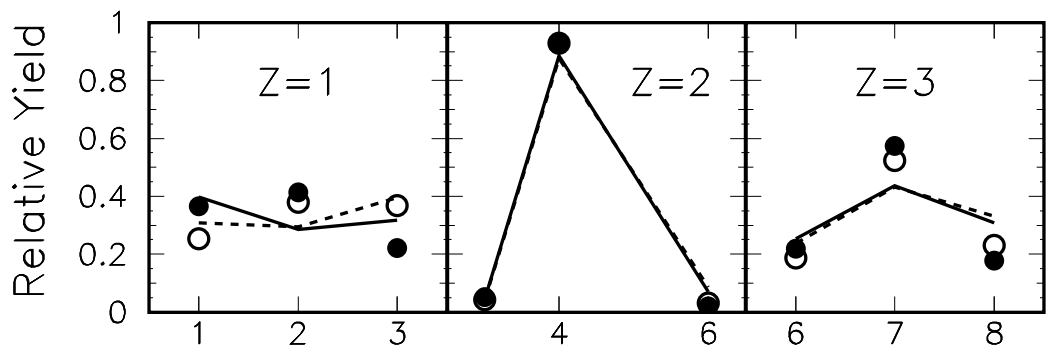
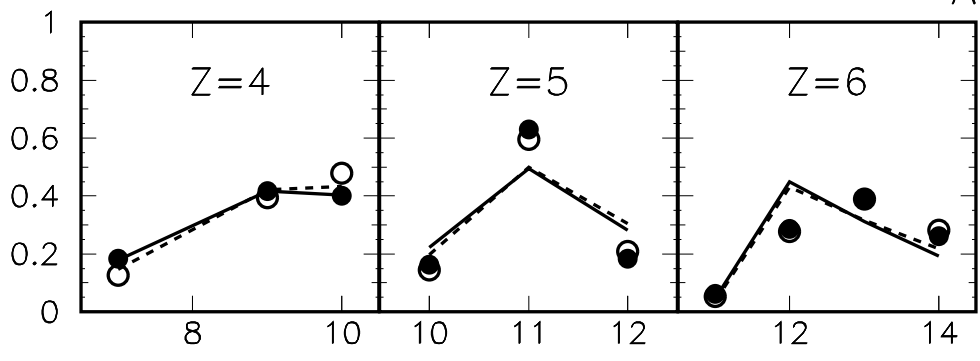
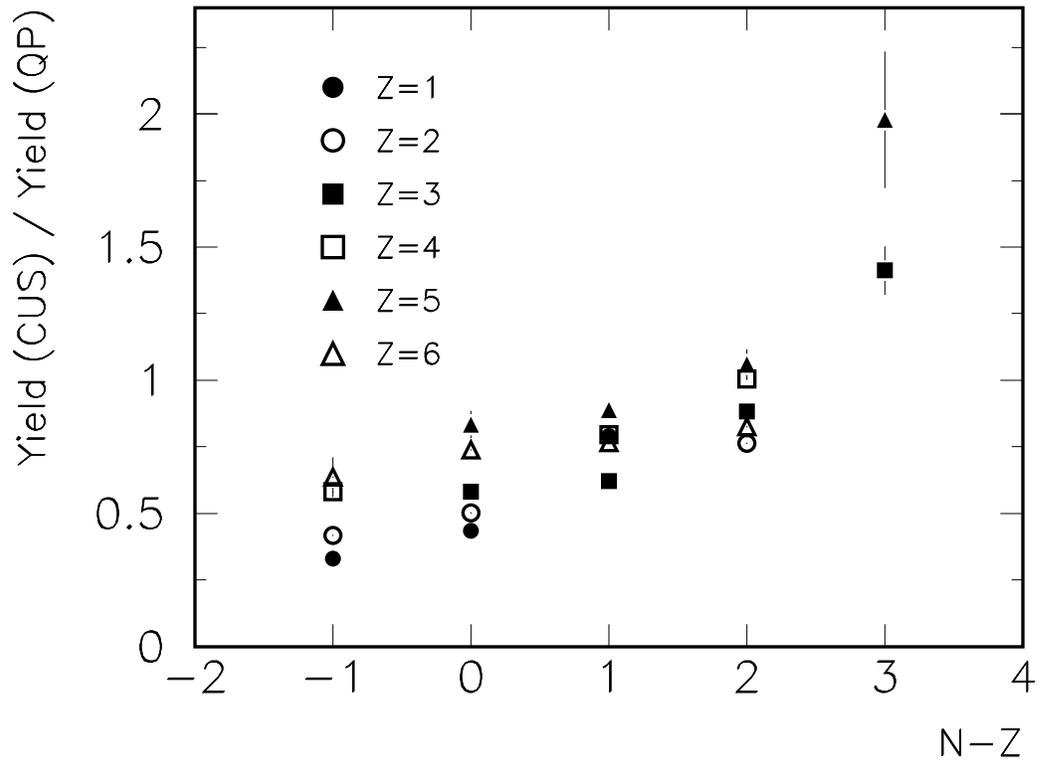
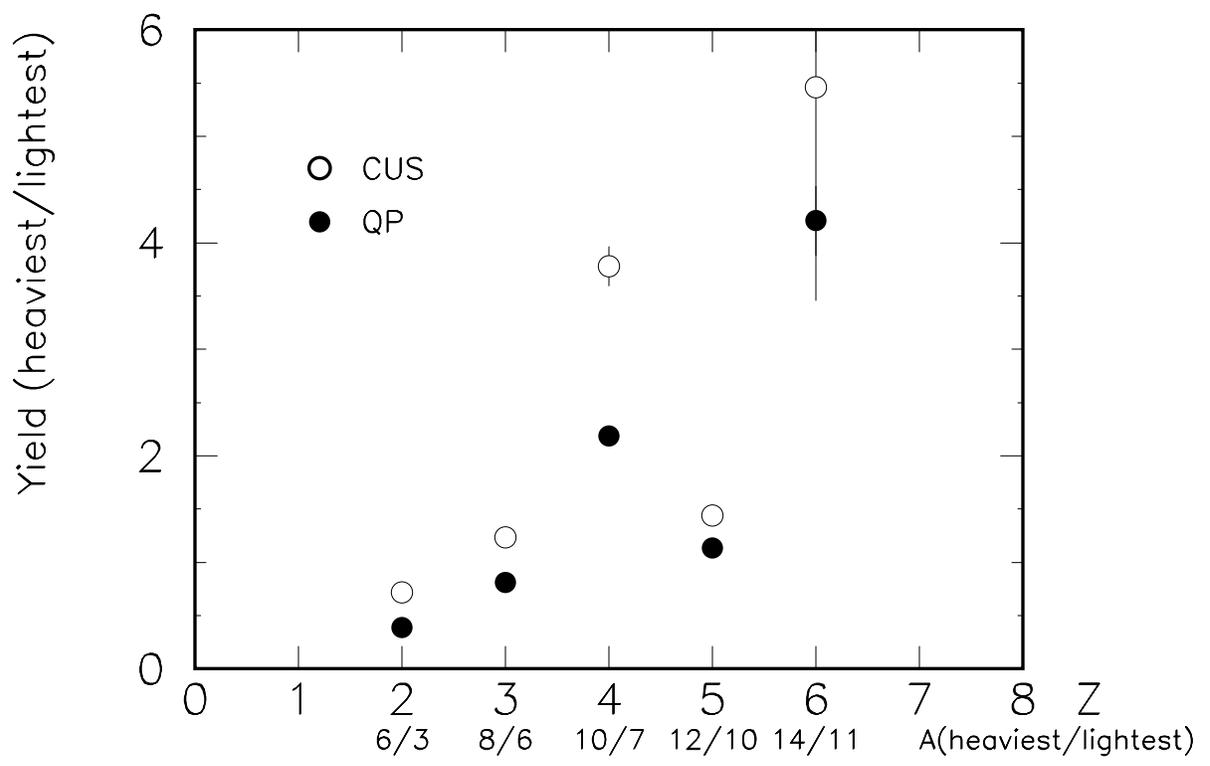
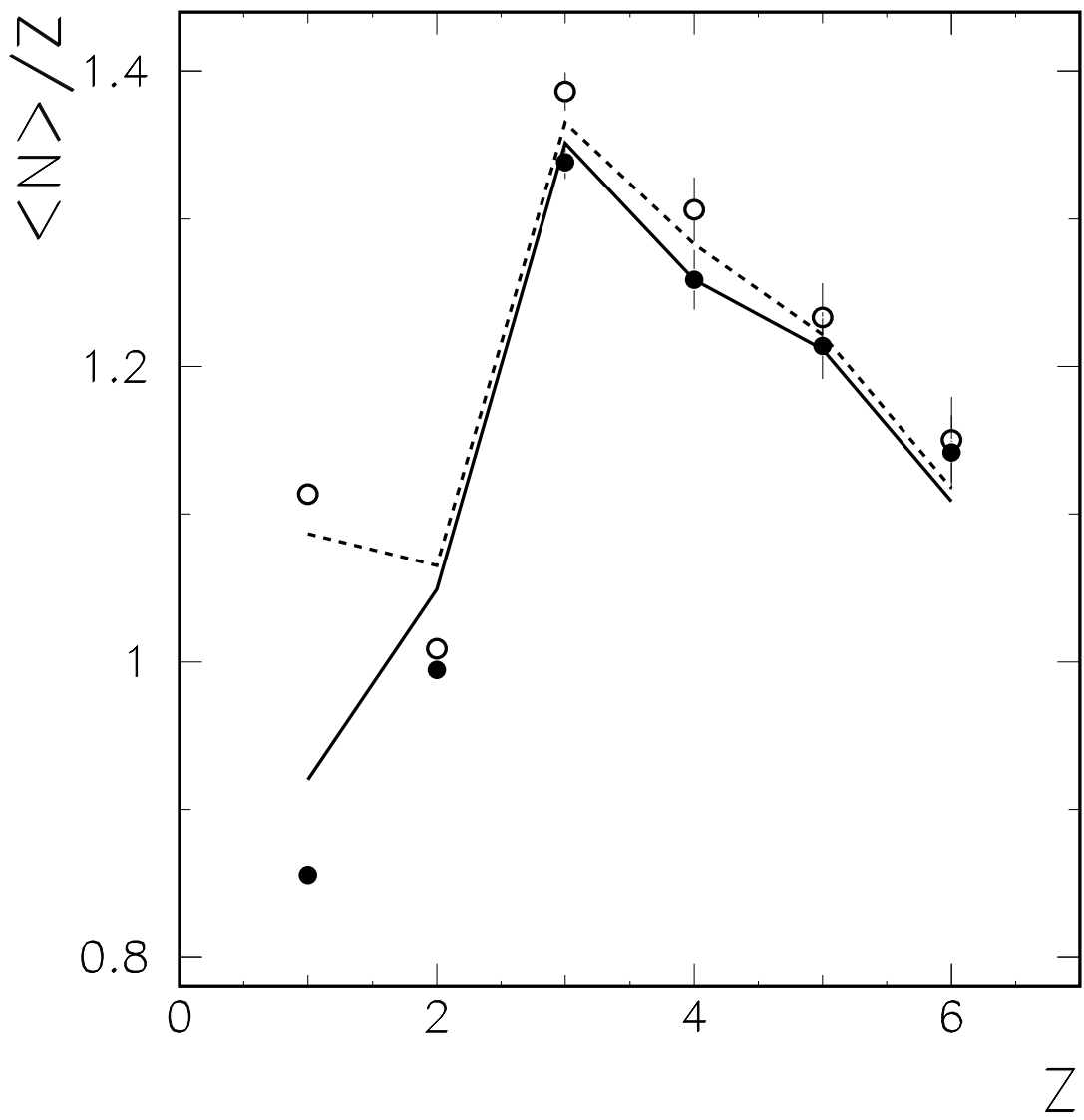
\end{document}